\newcommand{\bn}{{\mathbf n}}
\newcommand{\bv}{{\mathbf v}}

\newcommand{\cd}{\cdot}

\newcommand{\de}{\delta}

\newcommand{\be}{\begin{equation}}
\newcommand{\ee}{\end{equation}}

\newcommand{\bea}{\begin{eqnarray}}
\newcommand{\eea}{\end{eqnarray}}
\newcommand{\bean}{\begin{eqnarray*}}
\newcommand{\eean}{\end{eqnarray*}}

\newcommand{\apt}{{\ensuremath{\mathcal H}}}

\documentclass[prd,aps,floats,preprintnumbers,preprint]{revtex4-2}
\usepackage{graphicx,latexsym}
\makeindex
\begin{document}


\title{A discrete model of the effects of inhomogeneities on  the cosmological redshift}

\author{Antonio Enea Romano$^{1}$}
\affiliation
{
$^1$Instituto de Fisica, Universidad de Antioquia, A.A.1226, Medellin, Colombia\\
}

\texttt{}
\begin{abstract}
We develop a  discrete model  to study  the effects  of inhomogeneities on the redshift of photons propagating on cosmological scale. Inhomogeneities are modeled with independent discrete random variables at different positions along the propagation path, and the probability distribution of the cumulative effect on the redshift is computed using the moment generating function.
We find that the dispersion of the distribution is proportional to the square root of the comoving distance. 
We then discuss what are the  implications for the estimation of background cosmological parameters, giving some examples  for quantities related to the luminosity distance.
\end{abstract}

\begin{center}

\end{center}

\maketitle
\newpage
\section{Introduction}
Cosmology has some distinctive features respect to other branches of physics: we can observe the Universe only from one point, and we are unable to probe directly the geometry of space-time on large scales.
We can only detect particles or radiation which have reached us, and extract from these observations information about the space-time between us and the sources.
Most of cosmological scale distances can in fact only be deduced from redshift measurements, assuming an underlying space-time metric.
The Hubble law for example is based on the assumption of large scale isotropy and homogeneity of the metric describing the Universe. We know nevertheless that the Universe had inhomogeneities since very early times, as predicted by inflation, which are supposed to be the seeds for the later process of structure formation due to gravity.
There has been an long debate \cite{us,buchert,nambu,wald,tomita} on whether these perturbations of the homogeneous space-time metric could explain cosmological observations, in particular if they could induce effects equivalent to a cosmological constant or dark energy.
In this paper we consider what are the statistical effects of these inhomogeneities on the determination of background cosmological parameters, focusing on their complementary role rather than looking at them as alternatives to dark energy. It was shown in fact that even in presence of dark energy inhomogeneities could play an important role in affecting the determination of the background cosmological parameters \cite{Romano:2015iwa,Romano:2011mx,Romano:2014iea,EneaRomano:2011aa}.
In regard to the effects of inhomogeneities,  many observations suggest that the universe may be non-uniform, such as CMB, SN Ia, Galaxy cluster, GRB, FRB. In particular there is evidence of deviations from isotropy \cite{CosmoVerseNetwork:2025alb,Hu:2023eyf,Hu:2024big,Romano:2014iea}, which could reduce the Hubble tension.
The paper is organized as follows. We fist develop a discrete model of the inhomogeneities through which photons propagate from the source to the observer. We then use this model to estimate the effects of inhomogeneities on the determination of cosmological parameters and obtain the distribution probability function (PDF) for  the redshift correction, showing that its dispersion around the central value is proportional to the square root of the comoving distance. 
We then study the general implications on the estimation of background cosmological observables considering in particular the case of quantities related to the luminosity distance such as the Hubble parameter and the cosmological constant. 
\section{Effects of inhomogeneities on the luminosity distance}
Let's   consider a  Friedman universe with scalar perturbations. In
longitudinal gauge the metric can be written as 
\be\label{eq:conf}
ds^2 =  -(1+2\Psi)d\eta^2 +(1-2\Psi)\de_{ij}dx^idx^j ~,
\ee
where we have assumed assumed for simplicity the background to be flat and the perturbations to be of perfect fluid type.
Lets us suppose \cite{Sasaki,Dyer,Bonvin:2005ps} a photon is emitted from a source located at a comoving coordinate $r_S$ and time $t_S$ with a wave length $\lambda_S$, and is reaching an observer at the center $r_O=0$, with a wavelength $\lambda_O$ . Its total redshift can be expressed as the sum of two components . One is due to the homogeneous background expansion, $z_H$, and another due to the perturbations of the gravitation potential, $z_I$ :
\bea
\frac{\lambda_O}{\lambda_S}&=&1+z=1+z_H+z_I+z_{OS}=\frac{a(t_O)}{a(t_S)}( 1
+ \left[\psi +\bv\cdot \bn \right]^O_S -2\int_{\lambda_S}^{\lambda_O}d\lambda\dot\psi ~) ~\label{zpert},\\  z_H&=&\frac{a(t_O)}{a(t_S)}-1~,\\
z_I&=&(1+z_H)( -2\int_{\lambda_S}^{\lambda_O}d\lambda\dot\psi ~)~, \\
z_{OS}&=&(1+z_H)(\left[\psi +\bv\cdot \bn \right]^O_S ~)~, 
\eea

where we are denoting with a dot the derivative respect to conformal time $\eta$.

\section{Discrete approach } \label{Discrete}
In the following we will focus on the contribution $z_I$ coming from the integral, since we are interested in the cumulative effects of the structure between the source and observer. The other term $z_{OS}$ associated to the difference of the peculiar velocity and gravitational potential between the source and the observer is also important, but accurate observations of the source and information about our local Universe allow to determine it directly, making it a not  major source of uncertainty on the observed redshift, with  most of the  uncertainty coming from the  propagation of the errors of observationally estimated $\psi$ and $v$ of the source and the observer. See for example \cite{Riess:2016jrr} and references therein for an example of how density maps can be used to reconstruct the peculiar velocity and gravitational potential fields of the source using cosmological perturbation theory. For a non linear treatment see \cite{Vallejo:2017rga}.

On the contrary  the term $z_I$ associated to the integral is not under direct observational control because for a given source we do not exactly know what is the structure between it and the observer,  and the effects on the photons propagating through it are expected to increase with  the distance.

We can approximate $z_I$ discretely with
\bea
z_I&=&(1+z_H)(-2\int_{\lambda_S}^{\lambda_O}d\lambda\dot\psi ~)\approx -2(1+z_H)\Delta r\sum_{i=1}^N \dot\psi  ~,
\eea
where in the last approximate equality we have chosen the comoving coordinate as the affine parameter along the null geodesic, we have divided the interval $(r_S,0)$ into N subintervals of equal comoving length $\Delta r$ and approximated the integral with a sum. 

\begin{figure}
\begin{center}
\includegraphics[width=7cm,height=7cm]{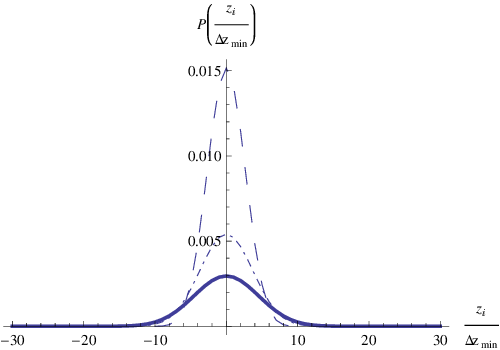}
\end{center}
 \caption{The probability distribution for $P(z_I/\Delta z_{min})$ is plotted for different values of N. 
 The thick line is for $N=30$, the dash-dotted line is for $N=20$, and the dashed line is for $N=10$. The difference between the different  distributions is due to
 the fact that as N increases $\Delta r$ decreases, lowering $\Delta z_{min}$. As it can be seen the dispersion increases with N.}
\label{fig:PS}
\end{figure}

The time evolution of the Fourier transform of the gravitational potential is given by:
\bea
\dot{\psi_k} &=&\psi_P(k)T(k)\frac{d}{d\eta}\bigg(\frac{D(\eta)}{a(\eta)}\bigg) ~,
\eea 
where $D(\eta)$ is the growth factor, $T(k)$ is the transfer function, and $\psi_P(k)$ is the primordial perturbation.
For a matter dominated universe the time derivative of the gravitational potential is zero, while for a Universe with cosmological constant there can be a significant contribution, which in the case of the cosmic microwave background radiation for example leads to the so called integrated ISW effect.

We can approximate the Gaussian distribution with a uniform discrete distribution for a variable $X$ with  three possible values 
\be
\{x_1,x_2,x_3\}=\{\sigma_{\psi},0,-\sigma_{\psi}\}. 
\ee

After substituting we get 
\bea
z_I&=&-2(1+z_H)\Delta r\sum_{i=1}^N \alpha_i \psi_i  ~, \label{eq:zid}
\eea
  where $\alpha_i$ is the value in the $i-th$ sub-interval of $\frac{d}{d\eta}\bigg(\frac{D(\eta)}{a(\eta)}\bigg)$ and
  $\psi_i$ is the value of the discrete random variable $X$  in the $i-th$ sub-interval determined by the time independent part of the $\psi(k,\eta)$, i.e. by $\psi_P(k)T(k)$. 
For large scales $T(k)$ is approximately 1, so $\psi_i$ is of the order of the primordial spectrum $\psi_P(k)$. 
	
	\begin{figure}
\begin{center}
 \includegraphics[width=7cm,height=7cm]{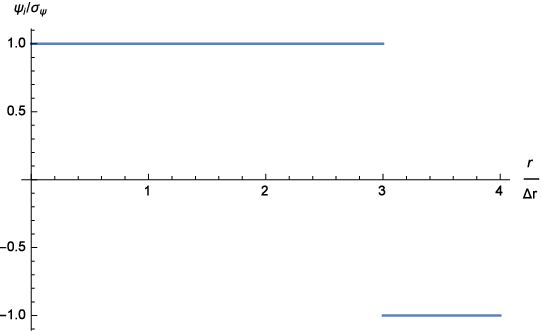}
\includegraphics[width=7cm,height=7cm]{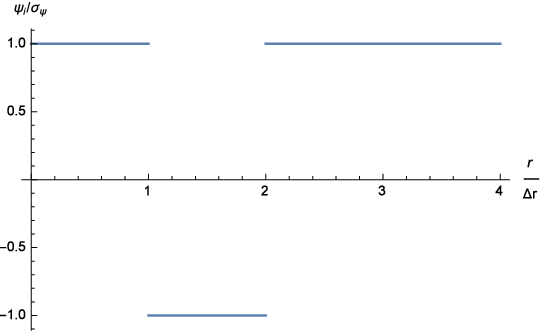}
\end{center}
 \caption{Two configurations of the discrete variable $\psi_i$ are shown as a function of the comoving distance. They both correspond to $\{n_{+}=3,n_{-}=1,S=2\}$ and consequently produce the same redshift correction $z_I$ as shown in eq.(\ref{zi}). This kind of degeneracy increases with the number of sub-intervals $N$ and implies a larger dispersion around the central value as shown in fig.(\ref{PS}).}
\end{figure}


\section{trinomial discrete approximation of a Gaussian process}
Assuming the primordial gravitational potential to be a treee dimesional Gaussian field, as predicted by many inflationary models and supported by cosmic microwave observations, also the gravitational potential along the propagation path will be a one-dimensional Gaussian field, as shown in section VI. We can then approximate it as a sequence of discrete random variables $\psi_i$, which can have three possible values: zero and plus or minus the standard deviation of the Gaussian field $\sigma_{\psi}$, with appropriate probabilities. The scaling limit of the sum of these variables is a trinomial discrete approximation of the Brownian motion, as implied by  the Donsker's theorem, also know as functional central limit theorem. In principle also a binomial approximation would suffice to get the right scaling limit as long as the variance and the mean are the same, but the advantage of the trinomial is that it can accommodate time varying variance. This is indeed modeling well the effect we are interesting in, since $\frac{d}{d\eta}\left(\frac{D}{a}\right)$ effectively changes the value of the standard deviation at different points along the path.

We have proposed a discrete model for the propagation of photons in an inhomogeneous medium. The large N limit of this discrete model has to coincide  with its continuos equivalent. We are modelling the potential along the photon path as a Gaussian process which can be approximated as a sequence of independent identically distributed random variables. This discrete model is valid for processes with constant power spectrum, i.e. white noise, and since the primordial curvature perturbations are approximately scale invariant, it should indeed be a good approximation.

The effect we are interested is the  integral of this Gaussian process, which is known to be a Browninan motion, and in fact our results will confirm this. A Gaussian process with constant spectrum  $\sigma^2$ can be approximated  as a sequence of random Gaussian variables with variance $\sigma^2/\Delta r$ \cite{sigmawhite}, where the interval over which the process is sampled discretely has been divided in $N=r/\Delta r$ sub-intervals. Under this discretization scheme in fact the  integral of the Gaussian process is
\bea
I&=&\int^r_0 X(r) dr \approx \sum^N_{i=1}\Delta r  X_i\,,
\eea
and since the variance of the a linear combination of a set of Gaussian variables  $X_i$ 
\be 
Z=\sum^N_{i=1} c_i X_i 
\ee 
has  variance 
\be
\sigma_Z^2=\sum^N_{i=1} c_i^2 \sigma_{X_i}^2\,,
\ee
from $\sigma_{X_i}^2=\sigma^2/\Delta r$ we get
\be
\sigma^2_I=N \Delta r^2 \frac{\sigma^2}{\Delta r}=N\Delta r \sigma^2=r\sigma^2\,.
\ee 
As expected in a Browonian motion, which is the  integral of a Gaussian process of the type we considered, the variance is linear in the time variable, which in this case is the conformal time, or equivalently since we are assuming a flat background, the comoving distance.


\section{Mean and variance of the variable s}
Putting all together we can model the integrand   $Y_i=\alpha_i\psi_i$  in each interval  with a uniformly distributed discrete random variable which can take three values 

\be
\{Y_{i,1},Y_{i,2},Y_{i,3}\}=\{\frac{\sigma\alpha_i}{\sqrt{\Delta r}},0,-\frac{\sigma\alpha_i}{\sqrt{\Delta r}}\}\,,
\ee
where we are using the notation $Y_{i,j}$ for the j-th value of the i-th variable and we are denoting $\sigma_{\psi}$ with $\sigma$. 
We can define the single variable moment generating function (MGF) for $Y_i$ as 
\bea
M_{\tilde{Y_i}}(t)&=&E[e^{t Y_{i}}]=\sum p_j e^{t Y_{i,j}}=\frac{1}{3}\left[1+\exp{\left(\frac{\sigma\alpha_i t}{\sqrt{\Delta r}}\right)}+\exp{\left(\frac{-\sigma\alpha_i t}{\sqrt{\Delta r}}\right)}\right] \,,
\eea
from  which we can get the mean and variance
\bea
\mu_{Y_i}&=&E[Y_i]=M_{Y_i}'(0)=0 \,, \\
\sigma^2_{Y_i}&=&Var[Y_i]=E[Y_i^2]-E[Y_i]^2=M_{Y_i}''(0)=\frac{2}{3}\frac{\sigma^2 \alpha_i^2}{\Delta r} \,. \label{sigmaYi}
\eea
From eq.(\ref{eq:zid}) we can see that the  effect we are interested in is proportional to the variable $S$ defined as
\bea
S&=&\sum^N_i \Delta r Y_i \,, \\
z_I&=&-2(1+z_H)\Delta r\sum_{i=1}^N Y_i=-2(1+z_H)\,S  ~. \label{zis}
\eea
Assuming $Y_i$ are mutually independent, i.e. as mentioned previously, modeling the Gaussian process with an approximately constant spectrum, the MGF for $S$ is given by the product of single variable MGF 
\bea
M_{S}(t)&=&\prod^N_i M_{Y_i}(\Delta r \,t) \,.
\eea
We can now compute mean and variance of $S$
\bea
\mu_S&=&E[S]=M_S'(0)=0 \,, \\
\sigma^2_{S}&=&Var[S]=E[S^2]-E[S]^2=M_S''(0)-M_S'(0)^2= \\
&=&\sum ^N_i \Delta r^2 \sigma_{Y_i}^2=\frac{2}{3} \sigma^2 \sum ^N_i \alpha_i^2 \Delta r \,.\nonumber \label{sigmas}
\eea
In the continuos limit we get
\bea
\sigma^2_{S}(z_H) &=& \frac{2}{3} \sigma^2 \int^{r(z_H)}_0 \alpha(r)^2 dr=\sigma^2 \beta(z_H) \,, \label{beta} \\
\beta(z_H)&=&  \frac{2}{3} \int^{r(z_H)}_0 \alpha(r)^2 dr \,.
\eea
Defining 
\bea
\overline{\alpha}^2&=&\frac{2}{3 N}\sum ^N_i \alpha_i^2=\frac{2}{3 N \Delta r}\sum ^N_i \alpha_i^2 \Delta r \label{alphac}\rightarrow \frac{2}{3 r(z_H)}\int^{r(z_H)}_0 \alpha^2(r) dr \,,
\eea
we can also express it as 
\bea
\sigma^2_{S}&=&N \Delta r \overline{\alpha} \sigma^2 = r \overline{\alpha}^2 \sigma^2 \,.
\eea


\section{Effects on the redshift}

Since the effect on $z_I$ is due to the integral along the line of sight of the time derivative of $\psi$, the parameter $\sigma_{\psi}$, which is the expected dispersion of the primordial $\psi$  averaged over intervals of length $\Delta r$,  can be computed using the relation between the one-dimensional (1D) and the three-dimensional(3D) power spectra \cite{1989MNRAS.238..293L}
\bea
P^{3D}&=&-\frac{2\pi}{k}\frac{d P^{1D}}{d k} \,.
\eea.
It turns out that the dispersion  computed from the 1D or 2D spectrum is the same \cite{Zaroubi:2005xx}
\bea
\sigma_{\psi}^2=\int d^3 {\bf{k}} P^{3D}_{\psi}=\int d k P^{1D}_{\psi}\,.
\eea
Observations of the cosmic microwave background radiation show that the 3D primordial power spectrum can be well approximated by a power law of the type
\bea
P^{3D}_{\psi} \propto k^{-3}k^n\,,
\eea
with $n\approx 0$. Note that this what is commonly called a scale invariant power spectrum, even if from a strictly mathematical point of view  $|\psi_k|^2$ depends on the scale even if $n=0$.

Assuming a scale invariant power spectrum of the form above and using a Gaussian filter to obtain a smooth  average over a scale $\Delta r$ the dispersion is
\bea
<\psi^2>_{\Delta r}&=\sigma^2_{\psi}(\Delta r)& \propto \int d^3 {\bf{k}} \, k^{n-3} e^{-k^2 (\Delta r)^2}  \propto (\Delta r)^{-n} \label{sigmapsi} \Gamma[\frac{n}{2}]\,.
\eea
Under the well observationally justified assumption of approximate scale invariance, i.e. for $n\approx 0$, we get that $\sigma^2_{\psi}(\Delta r)$ can be approximated as constant independent of $\Delta r$ which we will simply denote as $\sigma^2_{\psi}$. This is indeed the  characterizing  property of a Gaussian random field with a scale independent spectrum. 

We can now compute the dispersion of $z_i$ from eq.(\ref{sigmapsi}), eq.(\ref{sigmas}) and eq.(\ref{zis})

\bea
\sigma^2_{z_I}&=& [2(1+z_H) ]^2  \sigma^2_S = \frac{8}{3}   (1+z_H)^2 r \overline{\alpha}^2 \sigma^2  =\frac{8}{3}   \sigma^2 \beta(z_H) (1+z_H)^2\,. \label{szi}
\eea

\begin{figure}
\includegraphics[width=14cm,height=7cm]{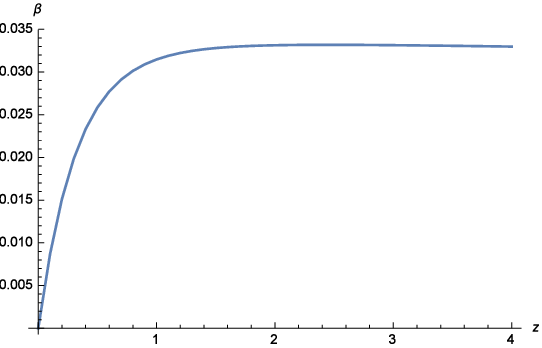}
\caption{The quantity $\beta$, as defined in eq.(\ref{sigmas}-\ref{beta}), is plotted as a function of the redshift in units of $H_0^{-1}$.  As it can be seen there is convergence towards an asymptotic value for large values of $z$ as expected from the fact $\alpha$ is asymptotically vanishing for large $z$. } \label{fig:beta}
\end{figure}

As expected the effect is independent of the number of subintervals $N$.
If we consider sufficiently large scales the CMB anisotropy observations give a good estimation of the parameter $\sigma^2 \approx 10^{-10}$, and since the asymptotic value of $\beta$ is about $0.035$ as shown in fig.(\ref{fig:beta}), we can conclude that the effect is of order unity only for very high redshift $z>10^5$, but even in that case the relative error $\sigma_{z_i}/z$ would be small and of order $10^{-5}$.

Note that according to eq.(\ref{alphac}) $\overline{\alpha}$ depends on $z_H$ and  is a decreasing function of the redshift.
This is due to the fact that $\alpha_i$ tends to zero at high redshift because the gravitational potential is frozen in a matter dominated Universe, so that the average through which $\alpha$ is defined in eq.(\ref{alphac}) is  dominated by the high redshift small values of $\alpha_i$.



\section{Behavior of $\overline{\alpha}$ }
In the continuous limit $\alpha_i$ is replaced by the  function
\bea
\alpha&=&\frac{d}{d\eta}\bigg(\frac{D(\eta)}{a(\eta)}\bigg)=\frac{d z}{d\eta}\frac{d}{d z}  \bigg(\frac{D(z)}{a(z)}\bigg)=-H(z)\frac{d}{d z }\bigg(\frac{D(z)}{a(z)}\bigg) \,.
\eea

For a flat matter dominated Universe, which should be a good model of the late time Universe in the epoch in which the time derivative of the gravitational potential is relevant, the growth factor is given by \cite{peebles:1993}
\bea
D(z)&=&E(z)G(z) \,,\\
G(z)&=& \frac{5 \Omega_m}{2}\int^{\infty}_z \frac{1+z'}{E(z')^3}d z' \,, \\
E(z)&=& \sqrt{\Omega_m(1+z)^3+1-\Omega_m} \,.
\eea

\begin{figure}
\includegraphics[width=14cm,height=7cm]{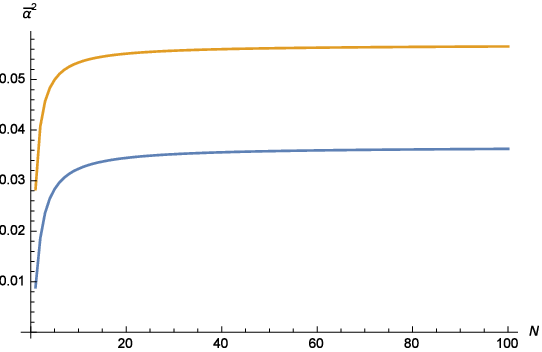}
\caption{The quantity $\overline{\alpha}^2$, as defined in eq.(\ref{alphac}), is plotted as a function of the number $N$ of intervals , in units of time of of $H_0^{-1}$. The blue curve is for $z_H=1$, and the orange curve is for $z_H=0.5$. As it can be seen there is convergence for large values of $N$. Also note that $\overline{\alpha}^2$ is smaller for larger values of $z_H$ because $\alpha$ is asymptotically tending to zero, making the average smaller.} \label{fig:alpha}
\end{figure}

As shown in fig.(\ref{fig:alpha}), $D/a$ is asymptotically constant during the matter domination phase, as expected from the freezing of the gravitational potential during that era.

\begin{figure}
\includegraphics[width=14cm,height=7cm]{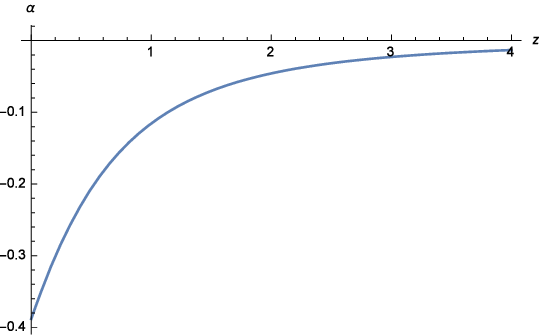}
\caption{The function $\alpha(z)$ is plotted as a function of the redshift, in units of time of of $H_0^{-1}$. As expected it tends to zero a high redshift, where the gravitational potential freezes.} \label{fig:alphac}
\end{figure}

\section{Probability distribution for the redshift correction}
From the above assumptions we can derive the probability distribution of  the redshift correction $z_I$, due to the spatial variations of the gravitational potential along the line of sight.
In this discrete model every space-time configuration along the photon path corresponds to assigning a set of N values to the random variable $\phi_i$.

Since as shown in eq.(\ref{alphac}) the variance $\sigma_{z_i}$ is the same for a set of different $\alpha_i$ or for a set of constant $\alpha_i=\overline{\alpha}$, we can assume that also the probability distribution could be well approximated under this assumption. Higher momenta could be different but this could be still a useful simplifying hypothesis to compute a probability distribution with the same variance as the one computed using the MGF method.

Let's call $\{n_+,n_-,n_0\}$, the number of positive, negative and zero values of the random variable $\psi_i$. The corresponding $z_I$ will be given by
\bea
z_I&=&-2(1+z_H)\overline{\alpha}\Delta r\sigma_{\psi}(n_+-n_-)=\Delta z_{min} S ~,\label{zi} \\ 
\Delta z_{min}&=&-2(1+z_H)\overline{\alpha}\Delta r\sigma_{\psi} ~,\\
S&=&(n_+-n_-) ~,
\eea 
where we have introduced $\Delta z_{min}$, the minimum possible inhomogeneous contribution to the total redshift.
According to our discrete model the total inhomogeneous redshift is a finite multiple of this minimum contribution, so we get:
\bea
-N \leq &\frac{z_I}{\Delta z_{min}}&\leq N ~,
\eea
where the lower and upper bounds correspond to configurations of all positive or all negative values of $\psi_i$, i.e. $n_+=N$ or $n_-=N$.
Since every interval can have three possible values of $\psi_i$, we have a total of $3^N$ possible different configurations.
Because of the fact that what it matters is the sum $(n_+-n_-)$, but not the order in which positive or negative perturbations appear, different space-time configurations can lead to the same value of $z_I$.
This degeneration is the origin of the intrinsic uncertainty on determining the origin of the total redshift, and consequently to estimate other background cosmological parameters such the $\Omega_{\Lambda}$.


The number of space-time configurations corresponding to a given value of $z_I$ can be calculated analytically.
Let's start from the number of configurations corresponding to a given set $\{n_-,n_+\}$  of positive and negative values of $\psi_i$, which is is given by:
\bea
N_c(n_-,n_+)&=&\frac{N!}{n_-!n_+!(N-n_--n_+)!} ~.
\eea
For a given $S$ the maximum $n_+$ corresponds to configurations with $n_0=0$, so from
\bea
n_++n_-+n_0&=&N \,,\\
n_+-n_-&=&S \,,
\eea
we get
\bea
n_+^{max}&=&Max[n_+,S]=\frac{N+S}{2} .
\eea 
The number of different configurations corresponding to a given value of $z_I$, i.e. to a given value of $S=(n_+-n_-)$, can be then computed  as
\bea
N_S&=&\sum^{}_{n_+-n_-=S}N_c(n_-,n_+)=\sum^{n_+^{Max}}_{n_+^{Min}}N_c(n_+-S,n_+)=\sum^{(N+S)/2}_{n_+=S}N_c(n_+-S,n_+)= \\
&& {}_2F_1\bigg[(1+S-N)/2;(S-N)/2;S+1;4\bigg]\frac{N!}{S! (N-S)!} \,,
\eea
where ${}_2F_1$ is the Gauss hypergeometric function.
It is easy to verify that due to the properties of ${}_2F_1$ $N_S$ is symmetric to respect to $S=0$, as expected from the symmetry of the uniform and symmetric probability distribution function $P(\psi)$ from which it is computed.

The formula derived above is consistent with the total number of possible configurations, i.e. we have
\bea
N_{tot}=\sum^N_{S=-N} N_S &=& 3^N ~.
\eea
The  PDF for S is finally given by 
\bea
P(S)=\frac{N_S}{N_{tot}}=3^{-N} {}_2F_1\bigg[(1+S-N)/2;(S-N)/2;S+1;4\bigg]\frac{N!}{S! (N-S)!} \label{PS}
\eea
which is plotted in fig.(\ref{fig:PS}) for different values of $N$, with $P(S)=P(\frac{z_I}{z_{min}})$.

\section{Effects on the estimation of background cosmological parameters}
We have seen that the dispersion of $z_I$ around the central value ($z_I=0$), is increasing as N increases, while its average value remains zero because of the symmetry of $P(S)$.
The estimation of background cosmological parameters requires to extract $z_H$ from the observed redshift $z_{obs}$, but due to our ignorance of the exact configuration of the gravitational potential along the line of sight, the term $z_I$ is effectively acting as an error for $z_H$, i.e.

\bea
z_H&=&z_{obs}-z_I \,,\\
\sigma_{z_H}^2&=&\sigma_{z_{obs}}^2+\sigma_{z_I}^2 \,, \label{szh}
\eea
where $\sigma_{z_{obs}}$ is the systematic observational error on the observed redshift $z_{obs}$.
This has a direct consequence on the estimation of any background cosmological parameter based on redshift observations. 
In general the measurement of any redshift dependent background observable quantity $Q(z_H)$ will have a propagated error
\bea
\sigma_Q^2(z_H)& \approx &\left(\frac{d Q}{d z}\right)^2\Big|_{z=z_{obs}}\sigma_{z_I}^2(z_{obs}) \,.
\eea
Since the function $\sigma_{z_I}^2(z_{obs})$ is growing monotonically because is proportional to the distance, the uncertainty $\sigma_Q(z_{obs})$ may grow or decrease as function of the redshift  depending on the behavior of the derivative $d Q/d z$.

\section{Effects on the luminosity distance}
In the following sections we will consider two specific cases related to the effects on the luminosity distance. 
Note that, following the arguments given in the beginning of section \ref{Discrete}, we are assuming that the effects on the redshift due to the difference of the peculiar velocity and gravitational potential between the source and the observer can be determined independently using other observations, and can be consequently subtracted  from the observed redshift, as done for example for supernovae in \cite{Riess:2016jrr}, a procedure called redshift correction in that context.

The effects of inhomogeneities on the luminosity distance $D_L(z)$ are not only due to the change in the redshift but also to  lensing and other integrated effects along the line of sight \cite{Sasaki,Dyer,Bonvin:2005ps}, but here will focus only on the effects due to $z_I$ since these are the ones  which we considered in this paper. 
For the other integrated effects terms, which are not due to the background redshift modification, a similar discrete approach could be adopted, but it would go beyond the scope of this paper, which is not the study of the effects of inhomogeneities on the luminosity distance, but more in general the study of the effects on the redshift and the consequences on cosmological observables estimation.
Other investigations of the effects of inhomogeneities on the luminosity distance using  non discrete approaches based on different averaging procedures applied to cosmological perturbations, including the effects we are not considering here, can be found for example in \cite{Fleury:2016fda,Ben-Dayan:2014swa,BenDayan:2013gc,BenDayan:2012ct}. 

Assuming a flat FRW background model the luminosity distance $D_L(z)$ and the comoving distance $r(z)$ are given by
\bea
r(z)&=&\int^z\frac{dz'}{H(z')} \,, \\
D_L(z)&=&(1+z) r(z) \,,
\eea
and the error on $D_L(z)$ due to $z_I$ is
\bea
\sigma_D^2& \approx &\left(\frac{d D_L}{dz}\right)^2 \sigma_{z_I}^2=\left[\frac{(1+z)}{H(z)}+r(z)\right]^2 \sigma_{z_I}^2\,.
\eea
As explained above the propagated error depends both on $\sigma_{z_I}$ and the derivative $D_L(z)$, which in this case are both monotonic growing  functions of the redshift. 


\section{Hubble parameter}
Let's consider for example the case  of the present value of the Hubble parameter $H_0$.
The local low redshift estimation of $H_0$ is based on the relation
\cite{Romano:2016utn}
\bea
H_0^{loc}&=& \lim_{z_H \to 0} \frac{z_H}{D_L(z_H)} \,. \label{H0loc}
\eea

According to the above expression the Gaussian propagated  error on $H_0^{loc}$ associated to the uncertainty of $z_H$ due to $z_I$ is
\bea
\frac{\sigma_{H_0}}{H_0} \approx \frac{1}{2} H_0 D_L''(z_{av})\sigma_{z_I}(z_{av}) \,.
\eea
where $z_{av}$ is the average redshift of the supernovae employed in the analysis.
In practice the value of $H_0$ is estimated using low redshift supernovae in a finite range \cite{Riess:2016jrr}, so the effect on the data fitting should be estimated more accurately using a procedure similar to the one given below. 

\section{Estimation of the cosmological constant}

Another interesting case could be the estimation of the cosmological constant using the luminosity distance of high redshift supernovae.
In this case the observational data could be binned in different redshift intervals and then for each one of them an error $\sigma_{z_I}$ could be estimated using eq.(\ref{szi}).
The $\chi^2$ variable which is minimized to fit the data is defined as
\bea
\chi^2&=&\sum^N_{j=1}\left[\frac{D^{obs}_L(z_j)-D^{th}_L(z_j)}{\sigma_{z_j}}\right]^2\,,
\eea
where we are denoting with $z_j$ the background redshift value $z_H$ of the j-th supernova and $\sigma_{z_j}$ is given in eq.(\ref{szh}).
The inclusion of the redshift dependent error given in eq.(\ref{szh}) in the data analysis is expected to have some effect on the estimation of $\Omega_{\Lambda}$ which only a careful analysis of observational data can quantify, and we leave it to a future work.

\section{Effects of pertubations correlation}
In the calculation of the probability distribution of the observed redshift we have assumed that the perturbations along the propagation path are independent, but large scale structure is not satisfying this simplifying assumption. This implies that the distribution computed analytically assuming a Brownian motion model does not fully capture the effects of realistic inhomogeneities on the redshift. In order to account for the effects of correlation between different sub-intervals a more realistic model should be used, such as a Markov Chain or an Autoregressive Process. 
For example we could model the correlation as
\be
Y_i = \rho \,Y_{i-1} + \epsilon_i
\ee
where $\rho = \exp(-\Delta r / L_c)$ is the lag-1 autocorrelation coefficient, $L_c$ is the characteristic correlation length of the gravitational potential, and $\epsilon_i$ is a white-noise  term.

The inclusion of the effects of correlation is expected to have an important impact on the redshift distribution, such as increasing the dispersion $\sigma_{z_I}$, increasing the error bounds, and inducing a transition from Brownian Motion to Super-Diffusion. We leave th study of these effects to a future work.
\section{Comparison to weak lensing effects on the lumonsity distance}
We have focused on the propagation of the inhomogeneity induced redshift uncertainty on observable quantities such as the luminosity distance, but  several other inhomogeneities effects can play an important role. At low redshift \cite{Bonvin:2005ps,Bolejko:2012uj} the dominating effect on the luminosity distance is in fact due to lensing. The redshift uncertainty contribution to the luminosity distance, which we have estimated by gaussian propagation of the redshift error, is in fact related to the integral of the time derivative of the gravitational potential, i.e. to the integrated integrated Sachs-Wolfe (ISW) effect, which is known to be subdominant with respect to the lensing effect \cite{Bolejko:2012uj}.
In this work we have focused on a discrete estimation of the redshift uncertainty due to the inhomogeneities effects along the propagation path, and then used this to obtain some estimation of its effects on observable quantities such as the luminosity distance, but in order to apply this to observational data, further study is required to include the effects of lensing, which could be achieved by employing a similar discrete approach to the density contrast along the propagation path.
\section{Full effects of inhomogeneities and general implications of redshift uncertainty}
In this work we have focused on the effects of inhomogeneities on redshift, due its fundamental importance in any cosmological observation. Nevertheless the effects of inhomogeneities on observable quantities can go beyond the propagation of the redshift uncertainty, since several other effects can be present, depending on the observable. For example, for the luminosity distance the leading order perturbations effects are given by \cite{Bonvin:2005ps} 
\begin{eqnarray}
	\tilde{d}_L(z_S,\bn)&=&(1+z_S)(\eta_O-\eta_S)\Bigg\{1
	-\frac{1}{(\eta_O-\eta_S)\apt_S}\bv_O\cd\bn
	-\left(1 -  \frac{1}{(\eta_O-\eta_S)\apt_S}\right)\bv_S\cdot\bn
	\nonumber\\ && 
	- \left(1- \frac{1}{(\eta_O-\eta_S)\apt_S}\right)\Psi_S  
	-\frac{1}{(\eta_O-\eta_S)\apt_S}\Psi_O
	 \hspace{-0.5cm}
	+ \frac{2}{(\eta_O-\eta_S)}\int_{\eta_S}^{\eta_O}d\eta \Psi \nonumber\\ &&
	+ \frac{2}{(\eta_O-\eta_S)\apt_S}\int_{\eta_S}^{\eta_O}d\eta \dot\Psi 
	-2\int_{\eta_S}^{\eta_O}d\eta
	\frac{(\eta-\eta_S)}{(\eta_O-\eta_S)}\dot\Psi 
	+\int_{\eta_S}^{\eta_O}d\eta
	\frac{(\eta-\eta_S)(\eta_O-\eta)}{(\eta_O-\eta_S)}\ddot\Psi
	\nonumber\\ &&  \quad
	-\int_{\eta_S}^{\eta_O}d\eta
	\frac{(\eta-\eta_S)(\eta_O-\eta)}{(\eta_O-\eta_S)}\nabla^2\Psi \Bigg\}
	~.  \label{eq:dL(zS)2}
\end{eqnarray}
where $\eta$ denotes the conformal time. This formula includes  the effects of the red-shift uncertainty propagation, related to the time derivative of the gravitational potential present in eq.(\ref{zpert}), and additional effects such as lensing, corresponding to the last term. 

While a full study of the inhomogeneities effects requires this kind of systematic approach, any observable with redshift dependence will include one contribution due to the propagation of the redshift uncertainty, so a general lower bound on the product of the redshift and observable $Q$ uncertainty can be obtained
\be
\sigma_Q^2(z_{obs}) \sigma_{z_I}^2(z_{obs}) \ge \left(\frac{\partial Q}{\partial z}\right)^2 \Big|_{z=z_{obs}} \sigma_{z_I}^4(z_{obs}) \,.
\ee
The actual lower bound could be higher, when including additional effects of homogeneities, such as for example the lensing effects for the luminosity distance, but it cannot be lower, since redshift uncertainty is always present.
The formula for the bound derived above is valid independently of the validity of the discrete approach adopted in this paper, i.e. the formula would still apply to the redshift uncertainty computed by correctly including the effects of the autocorrelation of the gravitational potential.

\section{Conclusions}
We have developed a  discrete model to account for the effects  of inhomogeneities on the redshift of photons. 
The model is based on assuming a discrete set values of the inhomogeneities in a given distance range, on summing all the contribution along the propagation path, and on computing the probability distribution of the sum of random discrete variables using the moment generating function.
Using this model we have computed the probability distribution of the redshift correction with respect to the background value, obtaining that its dispersion is proportional to the square root of the comoving distance. 
This dispersion is due to the fact that the observed redshift includes a contribution from the unknown configurations of the gravitational potential along the photon propagation path.
We then considered the implications for the estimation of background cosmological parameters, and giving some examples for quantities related to the luminosity distance.
The model reproduces discretely the expected stochastic properties of the propagation of photons in inhomogeneous media and  could be used for the numerical estimation of the effects of inhomogeneities on cosmological observables using Montecarlo methods.
In the future it will be interesting to compare the predictions based on this discrete approach to the results of numerical calculations for the propagation of photons in inhomogeneous media, and to include the effects of autocorrelation of the gravitational potential.

\section{Acknowledgments}
I thank Gabriele Veneziano for helpful discussions. 

\end{document}